\documentclass[aps,prd,showpacs,twocolumn]{revtex4}

\usepackage{epsfig}
\usepackage{longtable}
\usepackage{subfigure}
\usepackage{amsmath}

\begin{document}

\title{Using Final State Pseudorapidities to Improve $s$-channel Resonance Observables at the LHC}
\author{ Ross Diener$^1$, Stephen Godfrey$^{1,2}$\footnote{Email: godfrey@physics.carleton.ca} 
and Travis A. W. Martin$^1$\footnote{Email: tmartin@physics.carleton.ca}}
\affiliation{
$^1$Ottawa-Carleton Institute for Physics, 
Department of Physics, Carleton University, Ottawa, Canada K1S 5B6\\
$^2$TRIUMF, 4004 Wesbrook Mall, Vancouver BC Canada V6T 2A3 }

\date{\today}

\begin{abstract}

We study the use of final state particle pseudorapidity for measurements of $s$-channel 
resonances at the LHC.  Distinguishing the spin of an $s$-channel resonance can, in 
principle, 
be accomplished using angular distributions in the centre-of-mass 
frame, possibly using a centre-edge asymmetry measurement, $A_{CE}$. 
In addition, forward-backward asymmetry measurements,  $A_{FB}$, can be used to 
distinguish between models of extra neutral gauge bosons.  In this note
we show how these measurements can be improved by using simple methods 
based on the pseudorapidity of the final state particles 
and present the expected results for $A_{FB}$ and $A_{CE}$ for
several representative models.

\end{abstract}

\maketitle

\section{Introduction}
The startup of the CERN Large Hadron Collider (LHC) will allow 
the exploration of the TeV energy 
regime and the testing of the multitude of proposed theories of physics beyond the Standard Model. 
Many of these theories predict the existence of massive, neutral $s$-channel 
resonances~\cite{Hewett:1988xc,Langacker:2008yv,Rizzo:2006nw,Leike:1998wr,Cvetic:1995zs,
Randall:1999ee,Hewett:1998sn,Davoudiasl:2000wi,Kalinowski:1997bc,Kalinowski:1997zt}.
For some models of new neutral gauge bosons ($Z^{\prime}$), the LHC is expected to have
a discovery reach upwards of 5~TeV with 100~fb$^{-1}$ of integrated luminosity~\cite{Godfrey:1994qk}. 
This is a significant improvement over the current experimental limits for most models, which 
constrain $Z^{\prime}$ masses to values greater than 
$\sim 1$~TeV~\cite{CDF:2007sb,Erler:2009jh,Chivukula:2002ry}.

If a TeV scale $s$-channel resonance were discovered, the immediate task would be to identify its 
origins.   
Many observables have been proposed to this end, primarily focused on the 
dilepton channel ($e$ and $\mu$), which would produce 
the cleanest and most easily measured 
signal for a non-leptophobic 
$Z^{\prime}$~\cite{Hewett:1988xc,Langacker:2008yv,Rizzo:2006nw,Leike:1998wr,Cvetic:1995zs} 
with the ATLAS~\cite{:2008zzm,:1999fq} and CMS~\cite{Ball:2007zza} detectors. 
The proposed measurements for the dilepton channel are the $Z'$ width, total cross 
section, forward-backward asymmetry ($A_{FB}$)~\cite{Langacker:1984dc}, 
central-to-edge rapidity ratio~\cite{delAguila:1993ym}, 
and a comprehensive analysis of all rapidity regions~\cite{Petriello:2008zr,Li:2009xh}. 

There are challenges associated with some of these measurements that we argue can be 
alleviated by using the pseudorapidity of the final state fermions. 
In particular, we focus on two measurements: determining (or at least constraining) the spin of an 
$s$-channel resonance, and determining the forward-backward asymmetry of a $Z^{\prime}$.
Distinguishing whether the new resonance 
is a scalar, such as an R-parity violating sneutrino~\cite{Kalinowski:1997bc,Kalinowski:1997zt}, 
a spin-2 boson, such as a KK graviton~\cite{Randall:1999ee,Hewett:1998sn,Davoudiasl:2000wi}, 
or a spin-1 
$Z^{\prime}$~\cite{Hewett:1988xc,Langacker:2008yv,Rizzo:2006nw,Leike:1998wr,Cvetic:1995zs}
will be challenging and is typically determined through the study of the angular distribution in the 
centre-of-mass frame of the initial state quark and 
anti-quark (c.m.)~\cite{Davoudiasl:2000wi,Allanach:2000nr,Dvergsnes:2004tw,Osland:2008sy,Osland:2009tn}.
The forward-backward asymmetry measurement at the LHC has to deal with
the ambiguity in defining the forward
direction due to the inability to unambiguously determine the direction of the initial state
quark in a symmetric proton-proton collision.

Presently, some solutions exist to deal with these challenges.
To distinguish the spin of the resonance, a centre-edge asymmetry, 
$A_{CE}$,~\cite{Dvergsnes:2004tw} can be defined that is sensitive to the the angular 
distribution of the events.
The centre-edge asymmetry is a simple means of binning the events in the central and edge 
regions of $\cos\theta^{*}$, the c.m. scattering angle,
which will be weighted differently depending on the angular distribution.
This has the benefit of eliminating some of the systematic uncertainties of a fit to the angular 
distribution.
However, the $A_{CE}$ observable still relies on boosting the particle four-momentum from the lab 
frame to the c.m. frame.

The forward-backward ambiguity in a symmetric $pp$ collision can be resolved by exploiting the 
fact that the valence quarks have, on average, larger momentum than the sea anti-quarks.
The quark direction can then be identified with the boost direction of the dilepton 
system.~\cite{Dittmar:1996my}.
Restricting the measurement to those events that
have a large boost (i.e.: $|Y_{Z^{\prime}}| > 0.8$) 
reduces the misidentification of the initial
state quarks and antiquarks, 
resulting in greater than $70\%$ 
of dilepton events being correctly identified as being boosted by the quark~\cite{Dittmar:1996my}. 
Both of these methods have been explored in great detail 
and remain the standard approach used in the 
literature~\cite{Petriello:2008zr,Li:2009xh,Osland:2008sy,Osland:2009tn}.

Both the $A_{CE}$ and the $A_{FB}$ measurements 
require analysis of the centre-of-mass (c.m.) angular distribution of the 
dilepton events - directly for $A_{CE}$, and when tagging forward or 
backward events in $A_{FB}$.
In this note we propose a simpler method of measuring these asymmetries without 
reconstructing the angular distributions. 
Specifically, we exploit the direct measurements of the lepton pseudorapidities to 
calculate the observables, rather than using derived quantities that may propagate 
uncertainties into the result.
The proposed methods also take advantage of the fact that differences in pseudorapidities are 
Lorentz invariant quantities, so that all calculations can be performed using quantities 
measured in the lab frame.

The point of these methods is not to provide new phenomenological insight into the models, but 
rather to demonstrate how the use of final state pseudorapidities provides a simpler 
and cleaner means of obtaining the $A_{CE}$ and $A_{FB}$ values. 
The dimuon signal is very clean and error propagation should not be a big issue.
The real power of this approach will be seen when applied to heavy quark final states~\cite{Diener}. 
In the following sections, 
we give some calculational details which are followed by
a description of our approaches to the Centre-Edge Asymmetry and the Forward-Backward Asymmetry. 
We conclude with some final comments.

\section{Calculational Details}

The basic ingredients 
in our calculations are the cross sections,
$\sigma(pp\to R  \to \mu^+\mu^-)$, where $R=Z'$, $\tilde{\nu}$ or $G$.  
The cross section for $R=Z'$ is 
described by the Drell-Yan process with the addition of a 
$Z'$~\cite{Langacker:1984dc,Barger:1980ix,Godfrey:1994qk}.  
Analagous expressions for the spin-0 $\tilde{\nu}$ and spin-2 graviton are 
given in Ref.~\cite{Kalinowski:1997bc,Kalinowski:1997zt} and~\cite{Hewett:1998sn},
respectively. 
We computed the cross sections using Monte-Carlo phase space integration
with weighted events and imposed kinematic cuts 
to take into account detector acceptances, 
as described in the following sections.

In our numerical results we take $\alpha=1/128.9$, $\sin^2\theta_w=0.231$, $M_Z=91.188$~GeV, 
and $\Gamma_Z=2.495$~GeV~\cite{Amsler:2008zzb}.
We used the CTEQ6M parton 
distribution functions~\cite{Pumplin:2002vw} 
and included a K-factor to account for NLO QCD corrections \cite{KubarAndre:1978uy}. 
We neglected NNLO corrections, 
which are not numerically important
to our results~\cite{Melnikov:2006kv,Anastasiou:2003ds}, 
and final state QED radiation effects,
which are potentially important~\cite{Baur:2001ze} but require a detailed detector level simulation 
that is beyond the scope of the present analysis.
The $Z'$ widths only include decays to standard model fermions 
and include NLO QCD and electroweak radiative corrections \cite{Kataev:1992dg}.
For the $\tilde{\nu}$ width, we take $\Gamma_{\tilde{\nu}}=1$~GeV following 
Ref.~\cite{Kalinowski:1997zt}. Expressions for the $G$ width 
can be found in Ref.~\cite{Osland:2008sy,Allanach:2000nr,Han:1998sg,Bijnens:2001gh}.

\section{Spin Discrimination Using Centre-Edge Asymmetry, $A_{CE}$}

\begin{figure}[t]
   \begin{center}
   \leavevmode
   \mbox{}
   \epsfxsize=9cm
   \epsffile{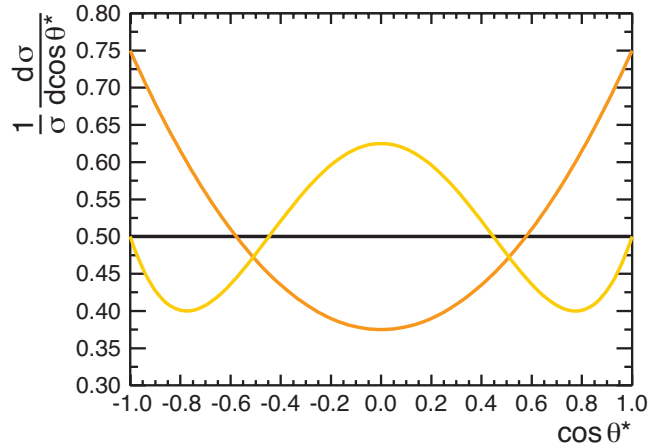}
   \end{center}
   \vspace{-10mm}
   \caption[ Normalized parton level angular distribution. ]
{\setlength{\baselineskip}{0.5cm} Normalized parton-level angular distribution of 
spin-0 (black), spin-1 (dark grey/orange) and spin-2 (light grey/yellow) bosons decaying to fermions.}
\label{fig:ang_z}
\end{figure}

The parton level angular distributions, $d\hat{\sigma}/d\cos\theta^{*}$, 
of the spin-0, -1, -2 bosons, shown in Fig.~\ref{fig:ang_z},
are distinct enough that, in principle, such a measurement would 
uniquely identify the spin~\cite{Davoudiasl:2000wi,Allanach:2000nr}. However, 
these distributions are not directly accessible due to the convolution with the parton 
distributions of the protons, the boosting of measured lab frame quantities to the 
centre-of-mass frame, detector limitations 
and finite statistics, all of which will make the 
measurement challenging~\cite{Allanach:2000nr}. 

The centre-edge asymmetry is almost entirely model-independent 
for spin-0 and spin-1 bosons.
For example, assuming the narrow width approximation for a 
$Z^{\prime}$, we find that $A_{CE} \approx 3/4\;\bar{z}(1+1/3\;\bar{z}^{2}) - 1/2$, 
for some value $\bar{z}$ that separates the centre and edge regions of $z^*$,
independent of the couplings to fermions. 
Spin-2 KK gravitons have contributions from 
$gg$ and $q\bar{q}$ processes that have slightly different angular distributions, 
and the $A_{CE}$ depends on the weighted contribution of each. 
The specific model will have an effect on the expected statistical uncertainties, but this 
should not be significant to the measurement due to the low backgrounds associated with 
leptonic final states.  
Thus, with limited statistics, an $A_{CE}$ measurement could have an advantage over a fit to 
the angular distribution.

\begin{figure}[t]
   \begin{center}
   \leavevmode
   \mbox{}
   \epsfxsize=9cm
   \epsffile{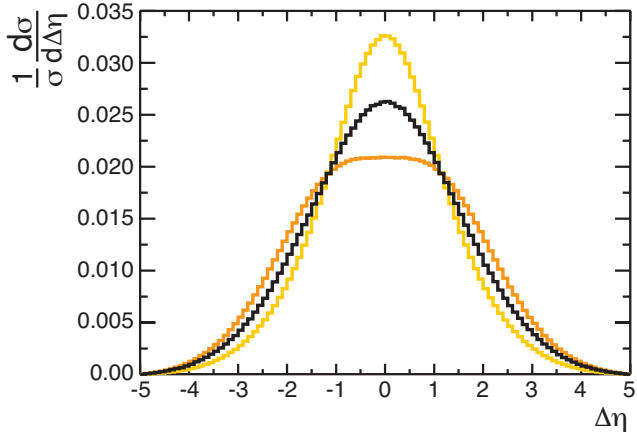}
   \end{center}
   \vspace{-10mm}
   \caption[ Normalized $\Delta\eta$ distribution including PDF and detector effects. ]
{\setlength{\baselineskip}{0.5cm} Normalized $\Delta\eta$ distribution including detector 
acceptance cuts ($|\eta_l| < 2.5$, $p_{T_l} > 20$~GeV) and only including events within 
$|M_{R} - M_{l^{+}l^{-}}| < 0.5\,\Delta M$. These cuts reduce the number of measurable 
events with large values of $|\Delta\eta|$.  $R = \tilde{\nu}$ (black), $Z^{\prime}$ 
(dark grey/orange), $G$ (light grey/yellow), where only one spin-1 distribution is shown due to the model independent nature of the spin-1 measurement. }
\label{fig:norm_de}
\end{figure}

For a hadron collider, the centre-of-mass angle of the outgoing fermion is not 
directly measurable on an event by event basis
due to the unknown values of the parton momentum fractions.
However, there exists a direct mapping between the c.m. angular distribution and the difference
in pseudorapidity of the final state lepton and anti-lepton, $\Delta\eta$.
Furthermore,  it is straightforward to show that $\Delta\eta$ is a Lorentz invariant quantity,
so that measuring this quantity in the lab frame is equivalent to measuring it in the
centre-of-mass frame: 
\begin{equation}
\Delta\eta_{lab} = \Delta\eta^{*} = \ln\left(\frac{1+\cos\theta^{*}}{1-\cos\theta^{*}}\right)
\label{eq:de_ct}
\end{equation}
The normalized $\Delta\eta_{lab}$ distributions for spin-0, -1, -2  
resonances are shown in Fig.~\ref{fig:norm_de}, 
where it is clear that they are distinct from one another.  
One can therefore construct a new centre-edge asymmetry using the lab frame 
$\Delta\eta_{lab}$ distribution
in place of the c.m. frame angular distribution.

Using the mapping given by Eq.~(\ref{eq:de_ct}), we define the centre-edge asymmetry:
\begin{equation}
\tilde{A}_{CE} 
= \frac{\left(\displaystyle\int^{\Pi}_{-\Pi} - \displaystyle\int^{-\Pi}_{-\infty} - 
\displaystyle\int^{\infty}_{\Pi}\right)\displaystyle\frac{d\sigma}{d\Delta\eta}d\Delta\eta}
{\displaystyle\int^{\infty}_{-\infty}\frac{d\sigma}{d\Delta\eta}d\Delta\eta}. 
\label{eq:ace}
\end{equation}
Following Osland, \textit{et al.}~\cite{Osland:2008sy}, we take $\bar{z} = 0.5$, 
which they find to be the ``optimal'' value, and translates to $\Pi = \Delta\eta = 1.099$. 
Experimentally, detector acceptance constrains the pseudorapidity of each fermion to 
$|\eta| < 2.5$, which limits $|\Delta\eta| < 5$ for the distribution.

As is common in the literature, we take an R-parity violating $\tilde{\nu}$ 
with $\lambda\lambda^{\prime} = (0.05)^{2}$ as an example of a spin-0 
resonance~\cite{Kalinowski:1997bc,Kalinowski:1997zt}, 
and an RS graviton with $c = 0.1$ for a spin-2 resonance \cite{Randall:1999ee}.
Current experimental 
limits and studies on model parameters are given in Ref.~\cite{Kalinowski:1997bc} 
and~\cite{Davoudiasl:1999jd}. 
For the $Z^{\prime}$ case we explored the 
$E_6$ models ($\psi$, $\chi$ and $\eta$)~\cite{Hewett:1988xc}, 
the Left-Right Symmetric model (LRSM, $g_{R} = g_{L}$)~\cite{Mohapatra:uf}, both 
the Littlest Higgs (LHM, $\tan\theta_{H} = 1.0$)~\cite{ArkaniHamed:2002qy} and 
Simplest Little Higgs (SLHM)~\cite{Schmaltz:2004de} models, 
and the Sequential Standard Model (SSM).

The spin-0 model, spin-2 and some $Z^{\prime}$ models we study predict 
narrow resonances, 
such that including events within several widths of the peak will be 
impossible in practice due to detector resolution effects smearing the 
Breit-Wigner distribution.
Instead, we examine events within one dilepton invariant mass bin as defined in the 
ATLAS TDR~\cite{:1999fq}, using $\Delta M = 42.9$~GeV for the 1.5~TeV resonance 
as in Ref.~\cite{Osland:2008sy}.
A more precise measurement could be obtained by including events from a wider invariant mass 
window, if a broader peak were to be observed.

In Table~\ref{table:ace} we show the expected centre-edge asymmetry for a spin-0, 
spin-1 and spin-2 resonance,
analogous to the study performed by Dvergsnes, \textit{et al.}~\cite{Dvergsnes:2004tw},
assuming muon final states with $96\%$ detection efficiency~\cite{:2008zzm}.
From Table~\ref{table:ace} one sees that if a $Z'$ were observed, 
a $G$ or $\tilde{\nu}$ could be ruled out.  
Likewise, an $A_{CE}$ measurement would strongly discriminate against
the $Z'$ or $\tilde{\nu}$ hypothesis if a $G$ were observed. 
However, the $Z'$ and $G$ hypothesis could only be ruled out at approximately $2.5\sigma$ 
if a $\tilde{\nu}$ signal was observed.
The primary limitation in distinguishing between the different possibilities is
the low statistics for $\tilde{\nu}$ production, as shown in the table, which is due 
to the tight constraints on the allowed values of its couplings.  
Other hypothetical spin-0 resonances may not be as tightly constrained and
could therefore be distinguished from a $Z'$ or $G$ with higher statistical significance.

\begin{table}[t]
\begin{center}
\caption[ Table of $A_{CE}$ values. ] {\setlength{\baselineskip}{0.5cm} 
$\tilde{A}_{CE}$ values with corresponding statistical uncertainties 
for 100~fb$^{-1}$ integrated luminosity, $p_{T_l} > 20$~GeV, $|\eta_l| < 2.5$, 
within one bin $\Delta M_{l^+l^-}=42.9$~GeV and $M_R=1.5$~TeV. 
Also shown are the expected number of total events 
for each model assuming 100~fb$^{-1}$ integrated luminosity.}
\vspace{2mm}
\begin{ruledtabular}
\begin{tabular}{ l l  r @{$\;\pm\;$} l l c }
\textbf{Model} & \;\;\;\;\;\; & $\tilde{A}_{CE}$ & $\delta \tilde{A}_{CE}$ & \;\;\;\;\;\; &\textbf{N Events} \\ \hline
$E_{6}\;\chi$ & &  $-0.106$ & 0.017 & & 3875\\ 
$E_{6}\;\psi$ & &  $-0.095$ & 0.022 & & 2223\\ 
$E_{6}\;\eta$ & &  $-0.092$ & 0.021 & & 2480\\ 
LR Symmetric & &  $-0.099$ & 0.018 & & 3350\\ 
Sequential SM & & $-0.097$ & 0.016 & & 4162\\ 
Littlest Higgs & &  $-0.095$ & 0.001 & & 6217\\ 
Simplest Little Higgs & & $-0.094$ & 0.017 & & 3542\\ 
RS Graviton & & $+0.228$ & 0.011 & & 8208\\ 
R-parity violating $\tilde{\nu}$ & & $+0.055$ & 0.066 & & 251
\end{tabular}
\end{ruledtabular}
\label{table:ace}
\end{center}
\end{table}

\section{Forward-Backward Asymmetry}

The forward-backward asymmetry is a well-established measurement for distinguishing 
between models of $Z^{\prime}$'s \cite{Langacker:1984dc}. 
For $p\bar{p}$ collisions at the Tevatron, 
the proton direction provides an obvious choice to define  the ``forward" direction. 
The choice of forward direction at the LHC is more subtle, 
and is conventionally defined as the direction of the $Z^{\prime}$ rapidity, 
$Y_{Z^{\prime}}$.
The $Z'$ rapidity is chosen because 
the parton distribution functions for the valence quarks peak at a 
higher momentum fraction than those of the anti-quarks, so the system has a higher
probability of being boosted in the quark direction.
This observation is statistical in nature and is more likely to hold true for larger values
of $|Y_{Z^{\prime}}|$. 
For smaller values of $|Y_{Z^{\prime}}|$, the momentum fractions 
of the quark and anti-quark are generally closer in magnitude,
so that using $|Y_{Z^{\prime}}|$ 
in the low rapidity region is less likely to correctly identify the quark direction.

\begin{figure}[t]
   \begin{center}
   \leavevmode
   \mbox{}
   \epsfxsize=9cm
   \epsffile{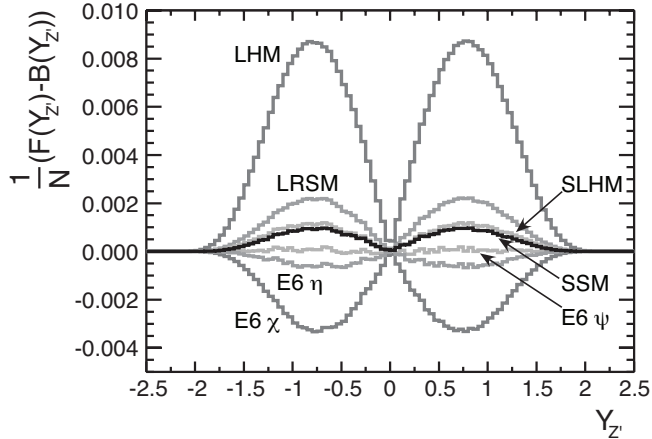}
   \end{center}
   \vspace{-10mm}
   \caption[ Rapidity distribution of difference between forward and backward events. ]
{\setlength{\baselineskip}{0.5cm} 
$A_{FB}$ as a function of the $Z'$ rapidity following Eq.~(\ref{eq:afb}) except that these
results are not integrated over rapidity.
From top to bottom, the models are 
LHM, LRSM, SLHM, SSM, $E_6\psi$, $E_6\eta$, $E_6\chi$.}
\label{fig:dsdy}
\end{figure}

A simpler method of defining a ``forward" or ``backward" event uses pseudorapidity. 
As before, we define the quark direction to be that of the higher momentum parton,  
or equivalently the direction of the $Z^{\prime}$ rapidity.
One can then show that a ``forward"
event is one in which $|\eta_{f}| > |\eta_{\bar{f}}|$ in the lab frame, 
and vice-versa for a ``backward" event.
Using these definitions for forward and backward, one can define the forward-backward asymmetry:
\begin{equation}
A_{FB} = \frac{\displaystyle\int \left[F(y) - B(y)\right] dy}{\displaystyle\int \left[F(y) + B(y)\right] dy}
\label{eq:afb}
\end{equation}
where $F(y)$ is the number of forward events and $B(y)$ is the number of backward events for a 
given 
$y$, the $Z^{\prime}$ rapidity (i.e.: $Y_{Z^{\prime}}$).
The $F(y)-B(y)$ distribution under this definition is clearly shown in 
Fig.~\ref{fig:dsdy} to be symmetric in $Z^{\prime}$ rapidity.
This method of finding $A_{FB}$ has the advantage of being very straightforward and clean.
It simply 
relies on counting events with $|\eta_{f}| > |\eta_{\bar{f}}|$ and those with
$|\eta_{f}| < |\eta_{\bar{f}}|$.
We note that a related technique is employed by the CDF collaboration ~\cite{Aaltonen:2008hc}
for the $Z^{0}$ $A_{FB}$ in $p\bar{p}$ collisions at the
Tevatron. However, the natural choice of the quark direction in $p\bar{p}$ 
collisions at the Tevatron in contrast to $pp$ collisions at the LHC 
results in important differences between the methods.

As in the conventional method for finding $A_{FB}$,
for small values of $|Y_{Z^{\prime}}|$,  
there is a higher probability to wrongly assume that the quark is the parton 
with the higher momentum fraction. This results in incorrectly assigning the forward or backward
direction and gives a small ``wrong'' contribution to the $A_{FB}$ measurement.
For this reason, it has been suggested that the central region, 
$|Y_{Z^{\prime}}| < Y_{min}$, be excluded in the measurement~\cite{Dittmar:1996my}.
However, the coupling dependency can still be determined without
this constraint on $|Y_{Z^{\prime}}|$ \cite{delAguila:1993ym}.

Another consideration for excluding the central region is that the number of events that remain
after subtracting $F-B$ is small, as shown in Fig. \ref{fig:dsdy}, 
while the total number of events in this region is large. 
Excluding the events in the central region
would increase the magnitude of $A_{FB}$, potentially making models more distinguishable. 
However, we found that increasing $Y_{min}$ resulted in an increase in the relative 
uncertainty.
We therefore conclude that little is gained by excluding events with small $Y_{Z^{\prime}}$, 
and suggest that the whole rapidity region be included to decrease uncertainty and further 
simplify the $A_{FB}$ measurement.

\begin{figure}[t]
   \begin{center}
   \leavevmode
   \mbox{}
   \epsfxsize=9cm
   \epsffile{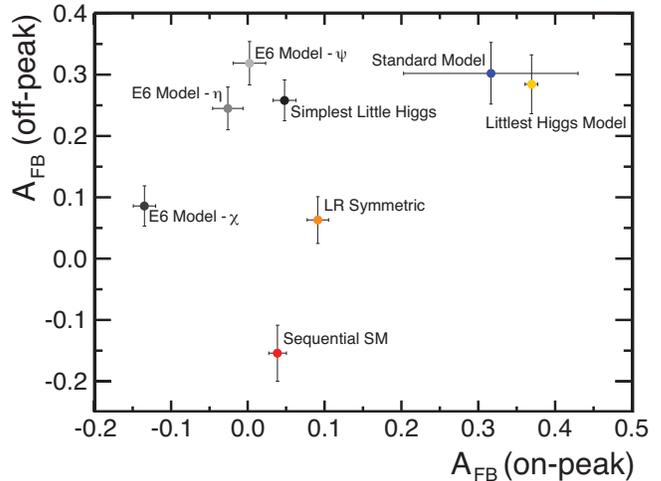}
   \end{center}
   \vspace{-10mm}
   \caption[ Forward-Backward Asymmetry off versus on-peak. ]
{\setlength{\baselineskip}{0.5cm} 
$A_{FB}$ off-peak versus on-peak for a variety of models, including detector acceptance limits
 and kinematic cuts as previously listed. Standard Model measurement determined from the standard 
model Drell-Yan cross section, with on-peak events within 
$|M_{l^{+}l^{-}} - M_{Z^{\prime}}| < 100$~GeV and off-peak events within 
$2/3 M_{Z^{\prime}} < M_{l^{+}l^{-}} < M_{Z^{\prime}} - 300$~GeV to include 
large enough statistics. }
\label{fig:AFB_on-off}
\end{figure}

Using this method, we calculate $A_{FB}$ for the $E_6$ models 
($\psi$,$\chi$,$\eta$)~\cite{Hewett:1988xc}, 
the Left Right Symmetric model~\cite{Mohapatra:uf}, 
the Littlest Higgs model~\cite{ArkaniHamed:2002qy}, 
the Simplest Little Higgs model~\cite{Schmaltz:2004de}, 
and the Sequential Standard model. The on-peak versus 
off-peak $A_{FB}$ are shown in Fig.~\ref{fig:AFB_on-off} , 
where on-peak includes events which satisfy 
$|M_{l^{+}l^{-}} - M_{Z^{\prime}}| < 3\Gamma_{Z^{\prime}}$ and off-peak includes events 
which satisfy $2/3 M_{Z^{\prime}} < M_{l^{+}l^{-}} < M_{Z^{\prime}} - 3\Gamma_{Z^{\prime}}$, 
similar to the cuts used by Petriello and Quackenbush \cite{Petriello:2008zr}.

We conclude with the important observation 
that it might also be possible to include some events in the 
forward regions of the calorimeter (FCAL) using this technique~\cite{Schram}. While a muon 
signature appears as missing $E_{T}$ in the FCAL, it may be possible to distinguish an electron 
from a jet in the FCAL due to the differences in the showering. 
The signal would require 
triggering off of a single, high $p_{T}$ electron in the $|\eta| < 2.5$ region, with an 
electron-jet in the FCAL. Determining the charge sign of the single electron would 
distinguish whether this is a ``forward" or ``backward" tagged event. It is not clear 
what the signal efficiency of this method is, as reducible backgrounds include $W+j$ and 
others that might have low rejection rates. Extending the rapidity range has
the potential of increasing the statistics and remains an interesting 
possibility for further study.

\section{Summary}

In this paper we described an approach for discriminating 
between various spin hypotheses 
for a newly discovered $s$-channel resonance
at the LHC using a centre-edge symmetry, $\tilde{A}_{CE}$, that is 
based on the difference
of the rapidities of final state fermions.  We also described a simple way to measure 
the forward-backward asymmetry, $A_{FB}$, 
using the properties of pseudorapidity. 
Both of these measurements have an advantage over previous approaches as they rely solely on the 
measurement of pseudorapidity, a fairly basic quantity. The new measurements require 
simple counting 
and should propagate fewer errors than previous approaches that rely on 
boosting the four momentum into the centre-of-mass frame in order to perform the analysis.  
Our approaches reproduce the results found in other analyses but via a more
straightforward analysis.

\acknowledgments

The authors would like to thank Malachi Schram and Isabel Trigger for helpful discussions.
This research was supported in part 
the Natural Sciences and Engineering Research Council of Canada.



\begin{thebibliography}{99}

\bibitem{Hewett:1988xc}
  J.~L.~Hewett and T.~G.~Rizzo,
  Phys.\ Rept.\  {\bf 183}, 193 (1989).

\bibitem{Langacker:2008yv}
  P.~Langacker,
  arXiv:0801.1345 [hep-ph].

\bibitem{Rizzo:2006nw}
  T.~G.~Rizzo,
  arXiv:hep-ph/0610104.

\bibitem{Leike:1998wr}
  A.~Leike,
  Phys.\ Rept.\  {\bf 317}, 143 (1999).

\bibitem{Cvetic:1995zs}
  M.~Cvetic and S.~Godfrey,
  arXiv:hep-ph/9504216.

\bibitem{Randall:1999ee}
  L.~Randall and R.~Sundrum,
  Phys.\ Rev.\ Lett.\  {\bf 83}, 3370 (1999)
  [arXiv:hep-ph/9905221].

\bibitem{Hewett:1998sn}
  J.~L.~Hewett,
  Phys.\ Rev.\ Lett.\  {\bf 82}, 4765 (1999)
  [arXiv:hep-ph/9811356].

\bibitem{Davoudiasl:2000wi}
  H.~Davoudiasl, J.~L.~Hewett and T.~G.~Rizzo,
  Phys.\ Rev.\  D {\bf 63}, 075004 (2001)
  [arXiv:hep-ph/0006041].

\bibitem{Kalinowski:1997bc}
  J.~Kalinowski, R.~Ruckl, H.~Spiesberger and P.~M.~Zerwas,
  Phys.\ Lett.\  B {\bf 406}, 314 (1997)
  [arXiv:hep-ph/9703436].

\bibitem{Kalinowski:1997zt}
  J.~Kalinowski, R.~Ruckl, H.~Spiesberger and P.~M.~Zerwas,
  Phys.\ Lett.\  B {\bf 414}, 297 (1997)
  [arXiv:hep-ph/9708272].

\bibitem{Godfrey:1994qk}
  S.~Godfrey,
  Phys.\ Rev.\  D {\bf 51}, 1402 (1995)
  [arXiv:hep-ph/9411237];
in {\it Proc. of the APS/DPF/DPB Summer Study on the Future of Particle Physics (Snowmass 2001) } ed. N.~Graf,
{\it In the Proceedings of APS / DPF / DPB Summer Study on the Future of Particle Physics (Snowmass 2001), Snowmass, Colorado, 30 Jun - 21 Jul
2001, pp P344}
  [arXiv:hep-ph/0201093].

\bibitem{CDF:2007sb}
  T.~Aaltonen {\it et al.}  [CDF Collaboration],
  Phys.\ Rev.\ Lett.\  {\bf 99}, 171802 (2007)
  [arXiv:0707.2524 [hep-ex]].

\bibitem{Erler:2009jh}
  J.~Erler, P.~Langacker, S.~Munir and E.~R.~Pena,
  arXiv:0906.2435 [hep-ph].

\bibitem{Chivukula:2002ry}
  R.~S.~Chivukula and E.~H.~Simmons,
  Phys.\ Rev.\  D {\bf 66}, 015006 (2002)
  [arXiv:hep-ph/0205064].

\bibitem{:2008zzm}
  G.~Aad {\it et al.}  [ATLAS Collaboration],
  JINST {\bf 3}, S08003 (2008).

\bibitem{:1999fq}
  ``ATLAS: Detector and physics performance technical design report. Volume
  1,'' CERN-LHCC-99-14
  ``ATLAS: Detector and physics performance technical design report. Volume
  2,'' CERN-LHCC-99-15

\bibitem{Ball:2007zza}
  G.~L.~Bayatian {\it et al.}  [CMS Collaboration],
  J.\ Phys.\ G {\bf 34} (2007) 995.

\bibitem{Langacker:1984dc}
  P.~Langacker, R.~W.~Robinett and J.~L.~Rosner,
  Phys.\ Rev.\  D {\bf 30}, 1470 (1984).

\bibitem{delAguila:1993ym}
  F.~del Aguila, M.~Cvetic and P.~Langacker,
  Phys.\ Rev.\  D {\bf 48}, 969 (1993)
  [arXiv:hep-ph/9303299].
  
\bibitem{Petriello:2008zr}
  F.~Petriello and S.~Quackenbush,
  Phys.\ Rev.\  D {\bf 77}, 115004 (2008)
  [arXiv:0801.4389 [hep-ph]].

\bibitem{Li:2009xh}
  Y.~Li, F.~Petriello and S.~Quackenbush,
  arXiv:0906.4132 [hep-ph].


\bibitem{Allanach:2000nr}
  B.~C.~Allanach, K.~Odagiri, M.~A.~Parker and B.~R.~Webber,
  JHEP {\bf 0009}, 019 (2000)
  [arXiv:hep-ph/0006114].

\bibitem{Dvergsnes:2004tw}
  E.~W.~Dvergsnes, P.~Osland, A.~A.~Pankov and N.~Paver,
  Phys.\ Rev.\  D {\bf 69}, 115001 (2004)
  [arXiv:hep-ph/0401199].
  
\bibitem{Osland:2008sy}
  P.~Osland, A.~A.~Pankov, N.~Paver and A.~V.~Tsytrinov,
  Phys.\ Rev.\  D {\bf 78}, 035008 (2008)
  [arXiv:0805.2734 [hep-ph]].
  
\bibitem{Osland:2009tn}
  P.~Osland, A.~A.~Pankov, A.~V.~Tsytrinov and N.~Paver,
  Phys.\ Rev.\  D {\bf 79}, 115021 (2009)
  [arXiv:0904.4857 [hep-ph]].
  
      
\bibitem{Dittmar:1996my}
  M.~Dittmar,
  Phys.\ Rev.\  D {\bf 55}, 161 (1997)
  [arXiv:hep-ex/9606002].

\bibitem{Diener}
R. Diener, S. Godfrey and T. Martin, in preparation.

\bibitem{Barger:1980ix}
  V.~D.~Barger, W.~Y.~Keung and E.~Ma,
  Phys.\ Rev.\  D {\bf 22}, 727 (1980);
  R.~W.~Robinett and J.~L.~Rosner,
  Phys.\ Rev.\  D {\bf 25}, 3036 (1982)
  [Erratum-ibid.\  D {\bf 27}, 679 (1983)];
  S.~Capstick and S.~Godfrey,
  Phys.\ Rev.\  D {\bf 37}, 2466 (1988).
See also Ref.~\cite{Hewett:1988xc,Langacker:2008yv,Rizzo:2006nw,Leike:1998wr,Cvetic:1995zs}
and references therein.

\bibitem{Amsler:2008zzb}
  C.~Amsler {\it et al.}  [Particle Data Group],
  Phys.\ Lett.\  B {\bf 667}, 1 (2008).

\bibitem{Pumplin:2002vw}
  J.~Pumplin {\it et al.}, 
  JHEP {\bf 0207}, 012 (2002).

\bibitem{KubarAndre:1978uy}
  J.~Kubar-Andre and F.~E.~Paige,
  Phys.\ Rev.\  D {\bf 19}, 221 (1979).

\bibitem{Melnikov:2006kv}
  K.~Melnikov and F.~Petriello,
  Phys.\ Rev.\  D {\bf 74}, 114017 (2006).

\bibitem{Anastasiou:2003ds}
  C.~Anastasiou {\it et al.}, 
  Phys.\ Rev.\  D {\bf 69}, 094008 (2004).

\bibitem{Baur:2001ze}
  U.~Baur {\it et al.}, 
  Phys.\ Rev.\  D {\bf 65}, 033007 (2002);
%
  U.~Baur, S.~Keller and W.~K.~Sakumoto,
  Phys.\ Rev.\  D {\bf 57}, 199 (1998);
%
  U.~Baur and D.~Wackeroth,
  Nucl.\ Phys.\ Proc.\ Suppl.\  {\bf 116}, 159 (2003).

\bibitem{Kataev:1992dg}
  A.~L.~Kataev,
  Phys.\ Lett.\  B {\bf 287}, 209 (1992).

\bibitem{Han:1998sg}
  T.~Han, J.~D.~Lykken and R.~J.~Zhang,
  Phys.\ Rev.\  D {\bf 59}, 105006 (1999)
  [arXiv:hep-ph/9811350].

\bibitem{Bijnens:2001gh}
  J.~Bijnens, P.~Eerola, M.~Maul, A.~Mansson and T.~Sjostrand,
  Phys.\ Lett.\  B {\bf 503}, 341 (2001)
  [arXiv:hep-ph/0101316].


\bibitem{Davoudiasl:1999jd}
  H.~Davoudiasl, J.~L.~Hewett and T.~G.~Rizzo,
  Phys.\ Rev.\ Lett.\  {\bf 84}, 2080 (2000)
  [arXiv:hep-ph/9909255].
  
\bibitem{Mohapatra:uf}
R.~N.~Mohapatra, {\em Unification And Supersymmetry. The
Frontiers Of Quark - Lepton Physics} (Springer, Berlin,
1986).
  
\bibitem{ArkaniHamed:2002qy}
  N.~Arkani-Hamed, A.~G.~Cohen, E.~Katz and A.~E.~Nelson,
  JHEP {\bf 0207}, 034 (2002)
  [arXiv:hep-ph/0206021].

\bibitem{Schmaltz:2004de}
  M.~Schmaltz,
  JHEP {\bf 0408}, 056 (2004)
  [arXiv:hep-ph/0407143].

\bibitem{Aaltonen:2008hc}
  T.~Aaltonen {\it et al.}  [CDF Collaboration],
  Phys.\ Rev.\ Lett.\  {\bf 101}, 202001 (2008)
  [arXiv:0806.2472 [hep-ex]].
  See also G.~Strycker, D.~Amidei, M.~Tecchio, T.~Schwarz, R.~Erbacher, J.~Conway, CDF Public Note 9724. [http://www-cdf.fnal.gov/physics/new/top/2009/tprop/Afb/]
  
\bibitem{Schram}
   M.~Schram, private communication.

\end{thebibliography}
\end{document}